# THE NETWORK ANALYSIS OF URBAN STREETS:
# A DUAL APPROACH

Sergio Porta[a], Paolo Crucitti[b], Vito Latora[c]

[a] Dipartimento di Progettazione dell'Architettura, Politecnico di Milano, Italy, sergio.porta@polimi.it

[b] Scuola Superiore di Catania, Italy, pacrucitti@ssc.unict.it

[c] Dipartimento di Fisica e Astronomia, Università di Catania,

and INFN Sezione di Catania, Italy, vito.latora@ct.infn.it

## Abstract

The application of the network approach to the urban case poses several questions in terms of how to deal with metric distances, what kind of graph representation to use, what kind of measures to investigate, how to deepen the correlation between measures of the structure *of* the network and measures of the dynamics *on* the network, what are the possible contributions from the GIS community. In this paper, the authors addresses a study of six cases of urban street networks characterised by different patterns and historical roots. The authors propose a representation of the street networks based firstly on a *primal graph*, where intersections are turned into nodes and streets into edges. In a second step, a *dual graph*, where streets are nodes and intersections are edges, is constructed by means of an innovative generalisation model named Intersection Continuity Negotiation, which allows to acknowledge the continuity of streets over a plurality of edges. Finally, the authors address a comparative study of some structural properties of the networks, seeking significant similarities among clusters of cases. A wide set of network analysis techniques are implemented over the dual graph: in particular the authors show that most of the considered networks have a broad degree distribution typical of scale-free networks and exhibit small-world properties as well.

# 1. Introduction

A large number of social, biological and man-made systems can be represented in the form of networks. For instance the society is made by individuals connected by social interactions (Wasserman and Faust, 1994), while the cell functioning is guaranteed by an intricate web of metabolites and chemical interactions. Equally, communication/transportation critical infrastructure systems, as the Internet (Pastor-Satorras and Vespignani, 2004) or a subway system (Latora and Marchiori, 2002; Arecchi et al. 2004) can be modelled as a network. The characterization of the topological properties of such networks has been the subject of a good deal of attention in the recent literature (Albert and Barabási, 2002). A variety of different variables have been proposed and thanks to the availability of powerful computers and large databases a huge number of real-world networks have been studied over the last few years.

The main result of this flurry of research in the scientific community has been that systems as diverse as the Internet, the actors' collaboration and the protein-protein interactions all share some common properties. In fact it has been shown that most of the studied networks exhibit the small-world property, meaning that in such networks the average topological distance between couples of nodes is small compared to the size of the network (it increases only logarithmically with the system size), despite the fact that the network has a large local clustering typical of regular lattices (Watts and Strogatz, 1998). Moreover it has been found that most real-world networks are scale-free, i.e. are characterized by the presence of hubs, nodes with a degree (a number of connections) k much larger than the average value $\bar{k}$. The empirical evidences collected from the analysis of both natural and man-created networks from the real world have shown in fact the presence of a power-law behaviour in the degree distribution $P(k) \sim k^{-\gamma}$ with the exponent γ varying between 2 and 3. The fact that most of the nodes have a small number of links, while a few have an extremely large number of connections, turns out to have extremely important consequences on the resilience of scale-free networks to errors and attacks. The emerging of scaling in a complex network has been recognized as the sign that the system is not static, but rather subject to incremental growth through time and preferential attachment (Albert and Barabási, 2002).

The network approach has been widely used in urban studies. Since the early sixties, a bulk of research has been spent trying to link the allocation of land uses to population growth through lines of transportation, or seeking the prediction of traffic flows given several topological and geometric characteristics of traffic channels, or eventually investigating the exchanges of goods and habits between settlements in the geographic space even in historical eras. Most if not all these approaches have been based on a quite simple, intuitive representation of networks which in short turns intersections (or settlements) into nodes and roads (or lines of relationship) into edges. The resulting graph is named in the context of this paper a primal graph. The primal approach took very soon the lead of network analysis implementations on territorial cases probably because it was the most simple way to capture one of the most crucial components of the geographic dimension: distance. As long as places are points and relations are edges, a value of distance can be easily associated to edges themselves, eventually interpreted as distance between places, which perfectly matches the ordinary, real-life experience of human beings.

On the other hand, an opposite representation happens to have sustained the by far most relevant, if not the sole, specific contribution of urban design to the study of city networks. After the seminal work of Hillier and Hanson (Hillier and Hanson, 1984) in the late eighties, Space Syntax has been developing a rather consistent application of the network approach to cities, neighbourhoods, streets and even single buildings, establishing a significant correlation between the topological accessibility of streets and phenomena as diverse as their popularity (pedestrian and vehicular flows), human way-finding, safety against micro-criminality, micro-economic vitality and social liveability (Hillier, 1996). Though not limited to "axial mapping", the core of the methodology is grounded on that particular process, through which the direct representation of a city plan, where intersections are nodes and streets are edges, is abandoned in favour of a dual representation, where streets are nodes and intersections are edges. More in detail, the axial map of a city pattern is a map where each straight space ("line of sight" or "line of unobstructed movement") is represented by one single straight line, an "axial line"; then, in the derivate syntax "connectivity graph", each axial line is turned into one node, while each intersection between any pair of axial lines is turned into one edge. At the end of the process, measures of accessibility (namely "integration") are calculated over the connectivity graph on the basis of a topological, non-Euclidean concept of distance (the so-called step-distance); finally, values of integration are represented back into qualified axial map layouts, which are the outcome of the analysis process. Space Syntax has been criticized for the largely subjective construction process of axial mapping (Jiang and Claramunt, 2002), its sensitivity to the edge-effect, as well as its difficulties to consistently explain some geometric configurations (Ratti, 2004), its distance to real life experience due to the abandonment of any reference to geographic-Euclidean space (Batty, 2004).

However, the advantage of the dual step-distance approach is one that can make the difference: because streets are mapped as nodes no matter their metric length, and because the intersections between every two streets are mapped as edges, one can have many – conceptually countless – intersections for each street, which means many – conceptually countless – edges for each node in the dual graph. This makes the dual graph of a geographic network comparable in its structure with most other networks recently investigated in social, biological and man-made systems, which in fact do not exhibit any geographic constrain. That leads, for instance, to the recognition of scale-free behaviour for the degree distribution of urban street networks, provided that they are represented with dual graphs (Rosvall et al. 2004).

## 2. Building the dual graph: the question of the generalization model and the Intersection Continuity Negotiation (ICN) proposal

A key question in the dual representation of street patterns is that a principle must be found that allows to extend the identity of a street over a plurality of edges in the primal graph; this problem, one of finding a generalization model, is about seeking a principle of continuity among different streets/edges, in order to capture the real sense of unity, or unique identity, of an urban street throughout a number of intersections. The question has been solved in Space Syntax substituting the primal graph representation of the network with the axial map – not properly a graph – where the principle of continuity is the linearity of the street spaces. After a first attempt to anchor the representation of street patterns to an actual primal graph, based on characteristic nodes and visibility (Jiang and Claramunt, 2002), Jiang and Claramunt have recently proposed one relevant model that builds a proper dual approach on a different primal representation (Jiang and Claramunt, 2004). Under their "named-street approach" the principle of continuity is the street name: two different arcs of the original street network are assigned the same street identity if they share the same street name. The main problem with this approach is that it introduces a nominalistic component in a pure spatial context, resulting in a loss of coherence of the process as a whole:

street names are not always meaningful in any sense, they are not always reliable as the same street may be termed in different ways by different social groups, or in different contexts, at different scales, in different ages. Other problems are that street name databases are not easily available for all cases or at all scales, and that the process of embedding and updating street names into GIS seems rather costly for large datasets.

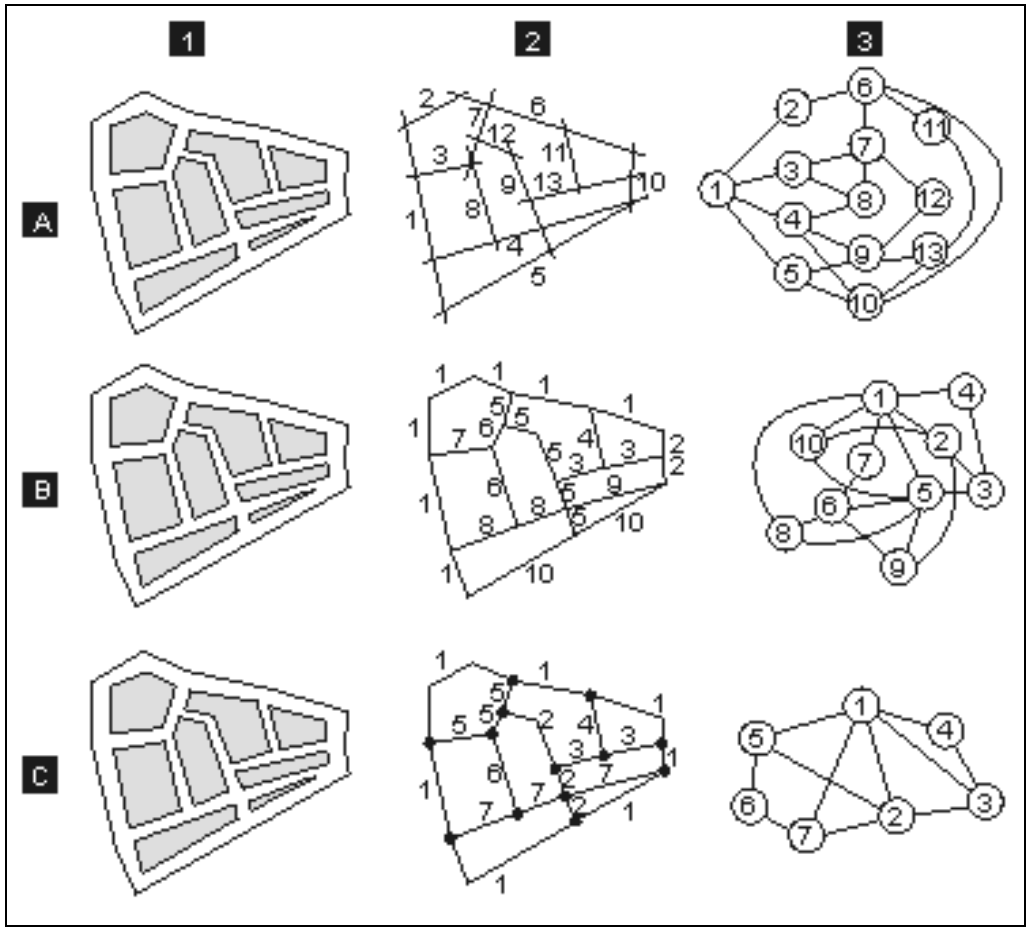

**Fig. 1**

Row A: the Space Syntax way: (1) A fictive urban system, its (2) primal axial map network model, and its (3) dual connectivity graph, after (Hillier and Hanson, 1984). Row B: the named street way (street names replaced by numbers): (1) A fictive urban system, its (2) primal network model, and its (3) dual connectivity graph after (Jiang and Claramunt, 2004). Row C: the proposed Intersection Continuity Negotiation (ICN) way (street names replaced by numbers): (1) A fictive urban system, its (2) primal graph, and its (3) dual connectivity graph. In this latter proposal, the direct representation of the urban network is properly a graph, where intersections are turned into nodes and street arcs into edges; edges follow the footprint of real mapped streets (a linear discontinuity does not generate a vertex); the ICN process assigns the continuity of street identities throughout any node.

However, implemented by Jiang and Claramunt on three real cases, the named-street approach has led to recognize a small-world character in large street networks, but no scale-free behaviour in their degree distribution.

In this work, we propose a new generalisation model of the primal graph based on a different principle of continuity (fig. 1). The model, which we term *Intersection Continuity Negotiation* (ICN), is quite simple and purely spatial, in that it excludes anything that cannot be derived by the sole geometric analysis of the primal graph itself. The model runs in three steps.

1. All the nodes are examined in turn. At each node, the continuity of street identity is negotiated among all pairs of incident edges: the two edges forming the largest convex angle are assigned the highest continuity and are coupled together; the two edge with the second largest convex angle are assignned the second largest continuity and are coupled together, and so forth; in nodes with an odd number of edges, the remaining edge is given the lowest continuity value.

2. Beginning with one edge chosen at random in the graph, a street ID code is assigned to the edge and, at relevant intersections, to the adjacent edges coupled in step 1.

3. The dual graph is constructed by mapping edges coded with the same street ID in the primal graph into nodes of the dual graph, and intersections among each pair of edges in the primal graph into edges connecting the corresponding nodes of the dual graph. Overlaying double edges in the dual graphs are eliminated.

The ICN generalisation model allows complex chains like loops and tailed loops to be recognized and nicely captures the most of the continuity of urban paths throughout urban networks. Being based on a primal graph, it minimizes subjectivity and re-enter the mainstream of the network representation of urban and territorial patterns. Being based on a pure spatial principle of continuity, it avoids problems of social interpretation within a pure spatial context. Finally, it allows a dual, step-distance representation of urban street networks linking it to a primal graph, which opens to further investigations in geographic-Euclidean space (Porta et al. 2004)

# 3. Characterizing the topological properties of a network

A network can be represented as a graph $G=(N, K)$, a mathematical entity defined by a pair of sets $N$ and $K$. The first set $N$ is a nonempty set of $N$ elements called *nodes* or *vertices*, while $K$ is a set of $K$ unordered pairs of different nodes called *links* or *edges*. In the following a vertex will be referred to by its order $i$ in the set $N$ ($1 \leq i \leq N$). If there is an edge between nodes $i$ and $j$, the edge being indicated as $(i,j)$, the two nodes are said to be adjacent or connected. A graph $G=(N, K)$ can be described by the adjacency matrix $A=\{a_{ij}\}$, a $NxN$ square matrix whose element $a_{ij}$ is equal to $1$ if $(i,j)$ belongs to $K$, and zero otherwise. In this section we present a list of the measures useful to characterize a graph and of the features that have been observed in real-world networks.

*3.1 Degree and degree distribution: scale-free networks*

The degree of a node is the number of edges incident with the node, i.e. the number of first neighbours of the node. The degree $k_i$ of node $i$ is defined as $k_i = \sum_{j \in G} a_{ij}$. The average degree is $\bar{k} = \frac{1}{N} \sum_{i \in G} k_i = \frac{2K}{N}$. Not all vertices in a network have the same number of edge. The way the degree is distributed among the nodes is an important property of a network that can be investigated by calculating the degree distribution $P(k)$, i.e. the probability of finding nodes with $k$ links. The degree distribution is defined as $P(k)=N(k)/N$, where $N(k)$ is the number of nodes with $k$ links. The study of a large number of complex systems, including man-made networks as the World Wide Web and the Internet (Pastor-Satorras and Vespignani, 2004), social networks, as the movie actors collaboration network or networks of sexual contacts (Liljeros et al. 2001), and biological networks (Albert and Barabasi, 2002), has shown that in most of the real systems the degree distribution follows a power law for large $k$:

$$P(k) \sim N(k) \sim k^{-\gamma} \qquad (1)$$

with the exponent γ being between 2 and 3. Networks with such a degree distribution are called scale-free (Albert and Barabasi, 2002). The results found are in contrast with what expected for random graphs (Erdös and Rényi, 1959). In fact, a random graph with N nodes and K edges (an average of $\bar{k}$ per node), i.e. a graph obtained by randomly selecting the K couples of nodes to be the connected, exhibits a Poisson degree distribution centred at $\bar{k}$.

*3.2 Degree correlations: assortative and disassortative mixing*

Another important quantity of a network is the correlation between the degree of connected vertices. In fact, it may happen either that high-degree nodes are preferentially attached to other high-degree nodes, or that they are connected to the low-degree ones. Both situations have been observed in real networks. The correlation between the degree of connected vertices can be quantified by considering $k_{nn}(k)$, i.e. the average degree of nearest neighbours of vertices with degree $k$ (Pastor-Satorras et al. 2001). Such a quantity is a constant as a function of $k$ if there are no correlations. If $k_{nn}(k)$ is an increasing function of $k$, vertices with low $k$ are connected to vertices with low $k$ and vertices with high $k$ are connected to vertices with high $k$. This property is referred in social science as assortative mixing, while a decreasing $k_{nn}(k)$ as a function of k is named disassortative mixing.

*3.3 Characteristic path length*

Social networks are historically the first complex networks explored. In one of the most famous experiments on social systems, Stanley Milgram asked a group of people, randomly selected in Omaha (Nebraska), to direct letters to a distant target person in Boston (Massachusetts). Letters had to be forwarded by an individual to a single personal acquaintance, thought to be closer to the final recipient. The experiment showed that the average number of steps from the sender to the final recipient, i.e. the acquaintance chain length, was only about six (Milgram, 1967). This phenomenon is often referred to as "six degrees of separation" (Guare, 1990). Analysis on other networks has shown similar properties: in most real-world networks it is possible to reach a node from another one, going through a number of edges that is small if compared to the total number of existing nodes in the system. The typical separation between two generic nodes in a graph $G$, can be measured by the characteristic path length $L$ defined as (Watts and Strogatz, 1998):

$$L(G) = \frac{1}{N \cdot (N-1)} \cdot \sum_{\substack{i,j \in G \\ i \neq j}} d_{ij} \qquad (2)$$

In this formula, $d_{ij}$ is the length of the shortest path between nodes *i* and *j*, i.e. the minimum number of edges covered in order to go from *i* to *j*.

### *3.4 Clustering coefficient*

Clustering is a property found in many real–world networks. For instance, in social systems, people show their inclination for self-organization in small communities within the system, and there is a high probability that two individuals linked by an acquaintance have a third acquaintance in common. Such tendency can be measured by calculating, as follows, the clustering coefficient *C* of a graph *G*. For each node *i*, we consider the subgraph $G_i$ of its first neighbours, that is obtained in two steps: 1) extracting *i* and its first neighbours from *G*; 2) removing the node *i* and all the incident edges. If node *i* has $k_i$ neighbours, then $G_i$ will have $k_i$ nodes and at most $k_i(k_i-1)/2$ edges. $C_i$ is proportional to the fraction of these edges that really exist, and measures the local group cohesiveness of vertex *i*. *C* is the average of $C_i$ calculated over all nodes:

$$C(G) = \frac{1}{N} \cdot \sum_{i \in G} C_i \qquad (3)$$

where

$$C_i = \frac{\# \text{ of edges in } G_i}{k_i \cdot (k_i - 1)/2} = \frac{\frac{1}{2} \sum_{l,m} a_{il} a_{lm} a_{mi}}{k_i \cdot (k_i - 1)/2} \qquad (4)$$

$C_i$ and consequently *C*, takes a value in the interval *[0,1]*. It is important to notice that *C* is related to the number of triangles present on the network (Latora and Marchiori, 2003). A more detailed description of the network can be obtained by plotting how $C_i$ is distributed among the nodes of the network (Jiang and Claramunt, 2004). For instance important information can be extracted by considering *C(k)*, the average clustering coefficient restricted to classes of vertices of degree *k*. In many cases *C(k)* exhibits a power law decay as a function of *k*, i.e. a hierarchy with low degree vertices belonging to well interconnected communities and hubs connecting many vertices not directly connected between each other. Various other measures to quantify the clustering of a group have been proposed over the years (Wasserman and Faust, 1994). Of particular relevance the generalization of the clustering coefficient *C* recently proposed to consider not only the immediate neighbouring nodes: the k-clustering coefficient $C^k$ measures to which extent the k-neighbours of a given node are interconnected with each other (Jiang and Claramunt, 2004).

*3.5 Global and local efficiency*

The global efficiency $E_{glob}$ is a measure of how well the nodes communicate over the network (Latora and Marchiori, 2001). The efficiency $\varepsilon_{ij}$ in the communication between node *i* and *j*, is assumed to be inversely proportional to the shortest path length, i.e. $\varepsilon_{ij}=1/d_{ij}$. In the case *G* is non-connected and there is no path linking *i* and *j* it is assumed $d_{ij}=+\infty$ and, consistently, $\varepsilon_{ij}=0$. The global efficiency of a graph *G* is defined as the average of $\varepsilon_{ij}$ over all the couples of nodes:

$$E_{glob}(G) = \frac{1}{N \cdot (N-1)} \cdot \sum_{\substack{i,j \in G \\ i \neq j}} \varepsilon_{ij} = \frac{1}{N \cdot (N-1)} \cdot \sum_{\substack{i,j \in G \\ i \neq j}} \frac{1}{d_{ij}} \qquad (5)$$

By definition $E_{glob}$ takes values in the interval [0,1], is equal to 1 for the complete graph, and is correlated to *1/L* (a high characteristic path length corresponds to a low efficiency). Consistently with the global analysis, we can measure the clustering properties of a graph by using the same measure, the efficiency, at the local level. The local efficiency is defined as (Latora and Marchiori, 2001):

$$E_{loc}(G) = \frac{1}{N} \cdot \sum_{i \in G} E(G_i) \qquad (6)$$

where

$$E(G_i) = \frac{1}{k_i \cdot (k_i - 1)} \cdot \sum_{\substack{l,m \in G \\ l \neq m}} \frac{1}{d'_{lm}} \qquad (7)$$

and $d'_{lm}$ is the shortest path length between node *l* and *m*, calculated in the subgraph $G_i$. A complex system can be therefore analyzed both in global and local scale by means of a single variable: the efficiency. As for $E_{glob}$, also $E_{loc}$ is already normalized for topological graphs.

*3.6 Small-world networks*

Random graphs have a characteristic path length *L* which is small with respect to the size of the system. In fact, it can be proven that $L = \log N / \log \bar{k}$, i.e. *L* grows only logarithmically with *N* (Erdös and Rényi, 1959). On the other hand, random graphs do not exhibit clustering: their clustering coefficient is $C = \bar{k}/N$ and tends to zero for large $N$. Conversely, a regular lattice has a finite clustering coefficient and a characteristic path length $L$ which grows linearly with the system size $N$. Watts and Strogatz have shown that many networks, ranging from social acquaintance networks, to networks in biology, have properties intermediate between random graphs and regular lattice. In fact, all such networks – that have been named small worlds – have at the same time: 1) a small average topological distance between couples of nodes, as random graphs; 2) a large local clustering, typical of regular lattices. To check whether a network is a small world it has been proposed to compare the value of $L$ and $C$ with those obtained for the randomized version of the network, i.e. for a network with the same $N$ and $K$ and in which the edges are distributed with a uniform probability among all the nodes (Watts and Strogatz, 1998). In a small-world network $L \sim L_{rand}$ and $C >> C_{rand}$.

An alternative definition of small worlds is based on the concept of network efficiency. Since a small characteristic path length indicates that the system is efficient on a global scale, and high clustering means that the network is efficient on a local scale, a small world is defined by having $E_{glob} \sim (E_{glob})_{rand}$ and $E_{loc} >> (E_{loc})_{rand}$, i.e. is a network in which the nodes communicate efficiently both at the global and at a local scale. It is important to notice that the notion of small-world can be given a more precise meaning concerning the global properties. In fact, the logarithmic scaling of $L$ as a function of the size is a precise way to characterize the small-world global property in growing network processes, where there is a meaningful range of systems sizes. More recently, Csanyi and Szendroi have proposed an alternative method valid for fixed networks as well. Denoting by $N_i(r)$ the number of nodes of the graph that can be reached from $i$ in at most $r$ steps, a small-world network will obey the scaling $N_i(r) \sim e^{\alpha r}$. So it is sufficient to check whether $\langle N(r) \rangle = \frac{1}{N} \sum_{i \in G} N_i(r)$, i.e. $N_i(r)$ averaged over all the nodes of the graph, grows linearly as a function of $r$ in a linear-log scale to prove that the network scale as a small world (Csanyi and Szendroi, 2004). Conversely, in many networks with strong geographical constraints, it has been found that $\langle N(r) \rangle \sim r^d$, which is the network discrete analogue of fractal scaling (Csanyi and Szendroi, 2004).

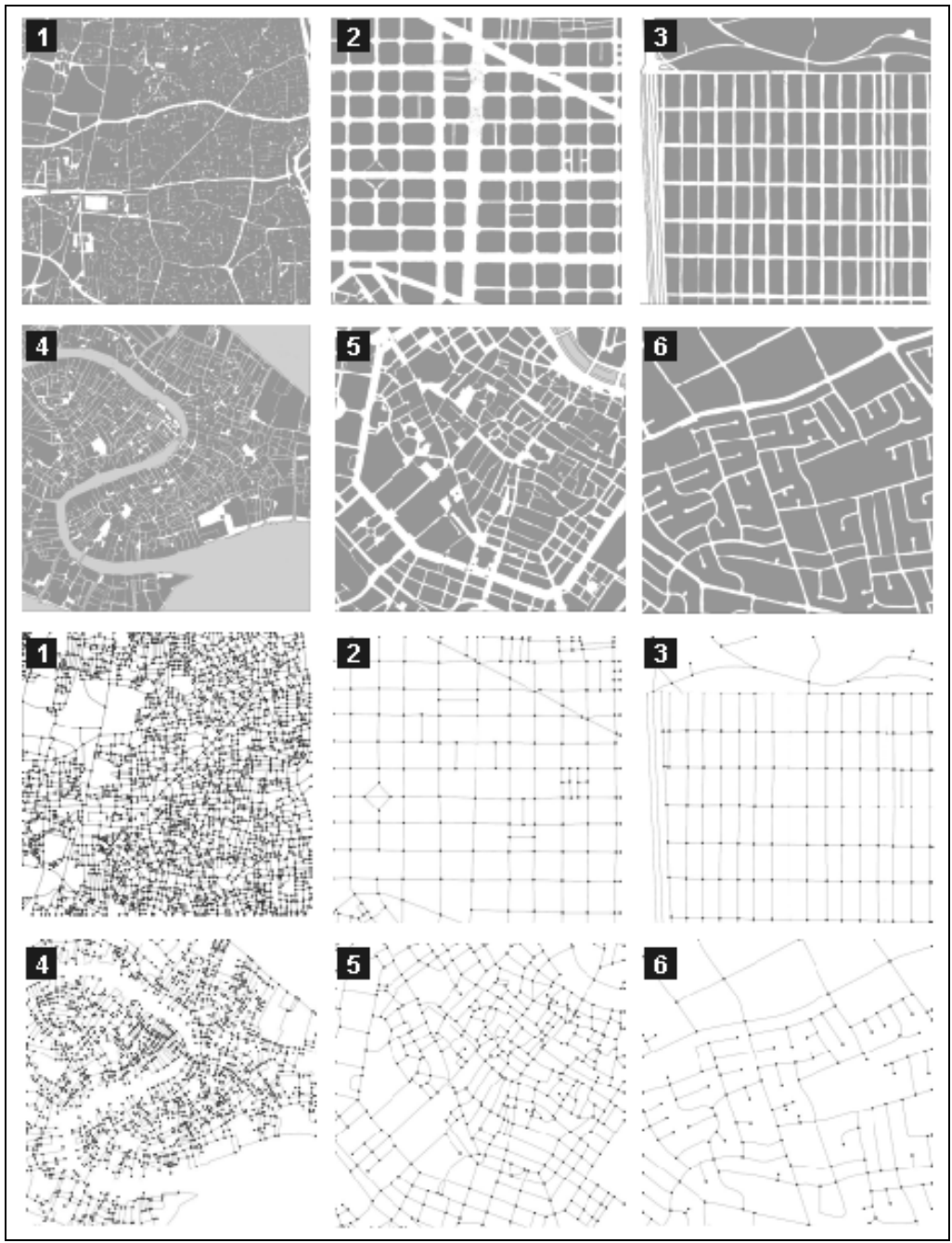

**Fig. 2**

The six 1-square mile samples of urban patterns (above) and their primal graphs (below): 1. Ahmedabad; 2. Barcelona; 3. San Francisco; 4. Venezia; 5. Wien; 6. Walnut Creek. Cities are so diverse that, at a first sight, it seems hard to imagine that they share any common – though hidden – pattern, which is what they actually do.

# 4. The 1-square mile project: comparative analysis of the topology of six urban street networks.

In this chapter we study some topological properties of six 1-square mile samples taken from different world cities. Drawing from a previous work of Allan Jacobs (Jacobs, 1993) we have selected six samples of urban patterns from different cities, also different in terms of structure, history and character; then we have imported them into GIS and firstly represented them as primal graphs (fig. 2, below); in so doing, we have turned intersections into nodes and streets into edges. Secondly, we have run the ICN generalisation model coding edges into generalized streets; thirdly we have developed the dual graphs mapping generalized streets as nodes and intersections as edges (fig. 3); finally, we have measured and compared the mentioned topological properties of the resulting dual graphs.

**Tab. 1**

The basic characteristic of the dual graphs obtained for the six 1-square mile urban patterns considered. We report the number of nodes $N$ (streets), the number of edges $K$ (intersections), the average number of edges per node $\bar{k}$, and the largest degree $k_{max}$.

| Case | $N$ | $K$ | $\bar{k}$ | $k_{max}$ |
|---|---|---|---|---|
| 1 Ahmedabad | 1239 | 2709 | 4.37 | 68 |
| 2 Barcelona | 53 | 168 | 6.34 | 15 |
| 3 San Francisco | 34 | 137 | 8.06 | 21 |
| 4 Venezia | 783 | 1312 | 3.35 | 29 |
| 5 Wien | 170 | 395 | 4.65 | 35 |
| 6 Walnut Creek | 78 | 107 | 2.74 | 13 |

Among cases, Ahmedabad, Venezia and Wien are historical, dense, mixed-use, windy fabrics originated by an incremental addition of urban materials across a long period of time, while Barcelona and San Francisco (grid-iron), and Walnut Creek ("lollipops") are modern patterns built in a relatively short period of time on the basis of one single plan. Thus, the selection of cases has been oriented to the discussion of the kind of order that emerges, though often not visible at a first glance, through an "organic" fine-grained growth out of the control of any central agency, as op-posed to the immediately visible Euclidean order showed in the case of most master-planned communities.

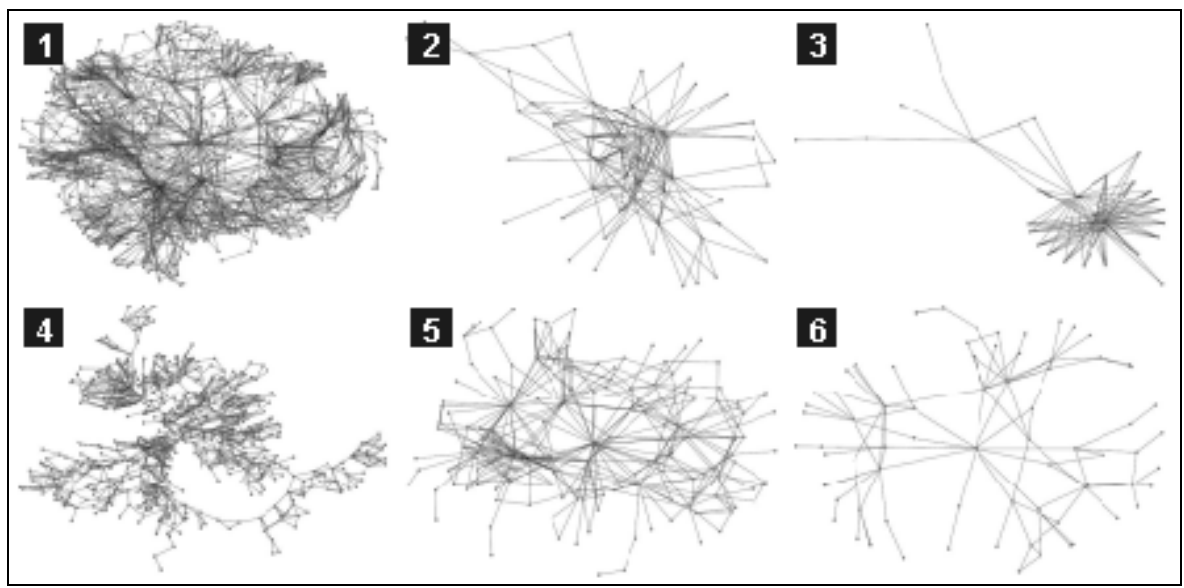

**Fig. 3**

The dual graphs of the six cities, shown in the same order of figure 2.

A first difference among cases is simply related to the number of streets and intersections. The dual graphs of cities are mapped in figure 3 while the number of streets (*N*) and intersections (*K*) are presented in table 1: for instance, Ahmedabad has *N*=1,235 and *K*= 2,705, while San Francisco has just *N*=34 and *K*= 137, meaning that in the same space of 1 square mile we found 1,235 streets in Ahmedabad and only 34 streets in San Francisco. On the other hand, if we compare $\bar{k}$, the average number of intersections per street (also reported in table 1), we find this value higher for Barcelona and San Francisco, respectively equal to about 6 and 8: that seems to be related to the grid-like structure of these two cities, which provides longer streets with more intersections. If we consider $k_{max}$ – the degree of the street with the highest number of intersections – we find that it is much larger than $\bar{k}$ in each of the six city considered, which is a clue of the large diversity of the nodes with respect to the number of intersections. In each case there are nodes with a small number of links but also a few nodes with an extremely large number of links. In some cases $k_{max}$ can be a consistent proportion of the total number of nodes of the graph: for instance in San Francisco (fig. 4), another grid-like case, we have found one street intersecting 21 other streets out of the 34 of the dual graph as a whole, a "degree coverage" of about the 62%.

The heterogeneity in the node degree can be better evidenced by plotting *N(k)* – the number of nodes with *k* links – as a function of *k* (see fig. 5). We have preferred to plot such a quantity instead of *P(k)* (see section 3.1) to remind the reader that the graphs considered have a wide differ-ent number of nodes *N*. All the reported distributions show the presence of long tails. A scale-free behaviour is clearly emerging in all graphs with a significant size, like Ahmedabad, Venezia and Wien. In the case of Ahmedabad, the graph with the largest number of nodes, we have fitted the distribution with a power law (the straight line reported in figure 5) extracting an exponent $\gamma = 2.5 \pm 0.1$

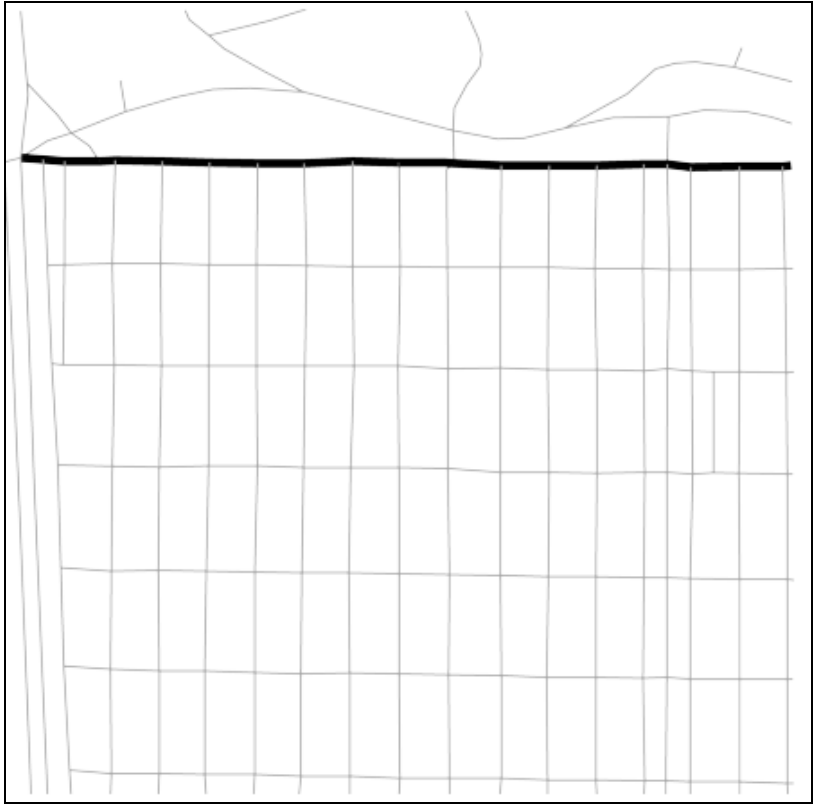

**Fig. 4**

The "richest" street (thick) of the case from San Francisco intersects 21 out of the 34 streets of the whole case. This corresponds to a "degree coverage" of a good 62% and is mainly due to the grid-like structure of San Francisco urban pattern.

The same distribution cannot be fitted by an exponential or a Poisson curve typical of random graph (see section 3.1). It should not be forgotten that all considered graphs are an expression of real street networks, all included in a 1 square mile boundary. Thus, though not meaningful in statistical terms, the small size of some graphs is here highly significant in urban terms, as it witness that some cities (i.e. the planned San Francisco, Barcelona and Walnut Creek) are patterned so that 1 square mile is simply too small to let any order emerge, while for others (namely the incrementally grown Ahmedabad, Venezia and Wien) the same "amount" of city is quite enough. That may tell a lot of a city, when issues of walkability, community cohesion and proxemic behaviours are at stake. Therefore, although a clear sign of the scale-free behaviour can be observed in large graphs only, there are a series of important indications that can be drawn from fig. 5 also for small cities. For instance, the peak in *N(k)* observed respectively for Barcelona around *k=12*, and for San Francisco at *k=7*, are fingerprints of the grid-like structure of these two urban patterns.

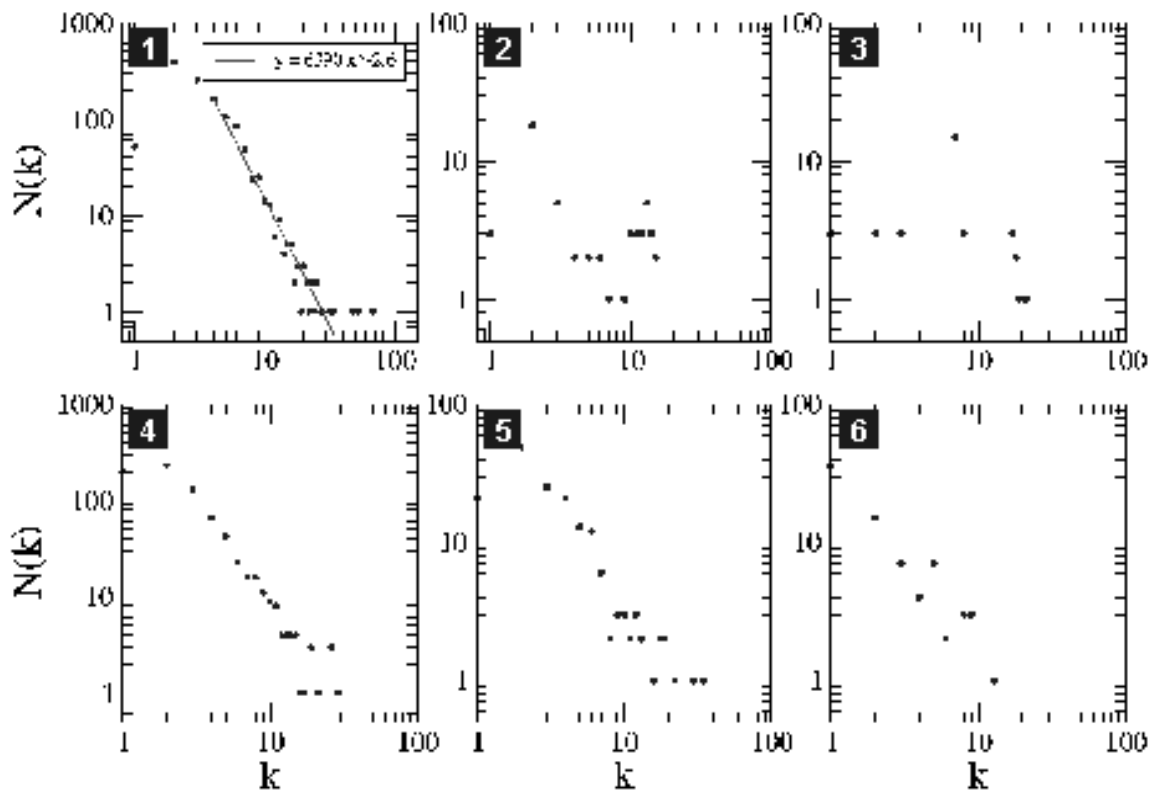

**Fig. 5**

Degree distribution of the six graphs. A scale-free behaviour is clearly emerging in graphs with a significant size like in the case of Ahmedabad: the fit reported has been obtained with a power law $N(k) \sim k^{-\gamma}$ with an exponent $\gamma = 2.5 \pm 0.1$. Since graphs are all an expression of real street networks, all included in a 1 square mile boundary, the small size of some graphs is here evident, so that in those cases (namely San Francisco, Barcelona and Walnut Creek) statistics does not allow to draw any precise conclusion on the presence or absence of a scale-free structure. Nevertheless in all the case considered the degree distribution are largely skewed.

A tendency toward a common, though not immediately evident order, clearly emerges in fine-graned, incrementally grown cities like Ahmedabad, Venezia and Wien, that correlates streets with their degree, thus the number of other streets intersected. Many streets intersect few other streets while a restricted number of "rich" streets do intersect a large number of other streets. In the case of Venezia for instance (fig. 6), within the upper 20 % interval of the degree range of values we find just 4 out of 783 streets (0.5 %, thick-black in the figure), while within the lower 20 % we find some 674 (86.1%, thin-grey).

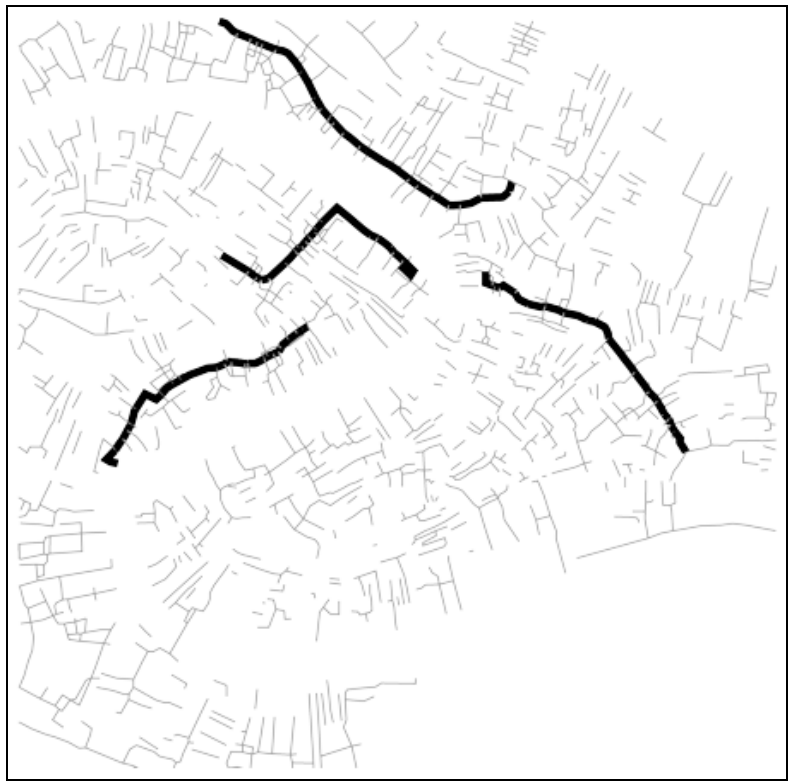

**Fig. 6**

Venezia is a good example of the disproportionate distribution of its 1,312 intersections across its 783 streets: in the interval of the upper 20% of intersections per street we find just 4 streets (thick, black lines), while in the lower 20% we find 674 (thin, grey lines), over the 85% of all streets. The distribution, however, is far from random: it clearly tends to a power law (see fig. 5).

But how rich are streets intersected by the richer? An important information can be obtained by plotting $k_{nn}(k)$, the average degree of nearest neighbours of vertices with degree $k$. As discussed in Section 3.2 such a plot can tell us if there are correlations between the degree of connected vertices. In fig. 7 we observe that both Venezia and Wien show a visible tendency to disas-sortativity. In general, assortative mixing, that is typical of many social systems, is not detected in any of the six urban networks here considered. Such a result is probably related to a principle of hierarchy which drives rich streets to "order" the urban pattern at the local level: to have many rich streets intersecting each other would lead to a waste of land and financial resources, for one single "main street" can easily and rather successfully connect the most of an urban district.

The only notable exception to this rule seems to be San Francisco, a case in which we observe, for $k$ smaller than 7, an increasing $k_{nn}$ as a function of $k$. In particular, the peak at $k=7$ is due to the fact that in the 1-square mile sample of San Francesco there is a large number of streets with $k=7$, namely the vertical streets, all of them intersecting the horizontal street with the largest degree $k=21$ (see fig. 4).

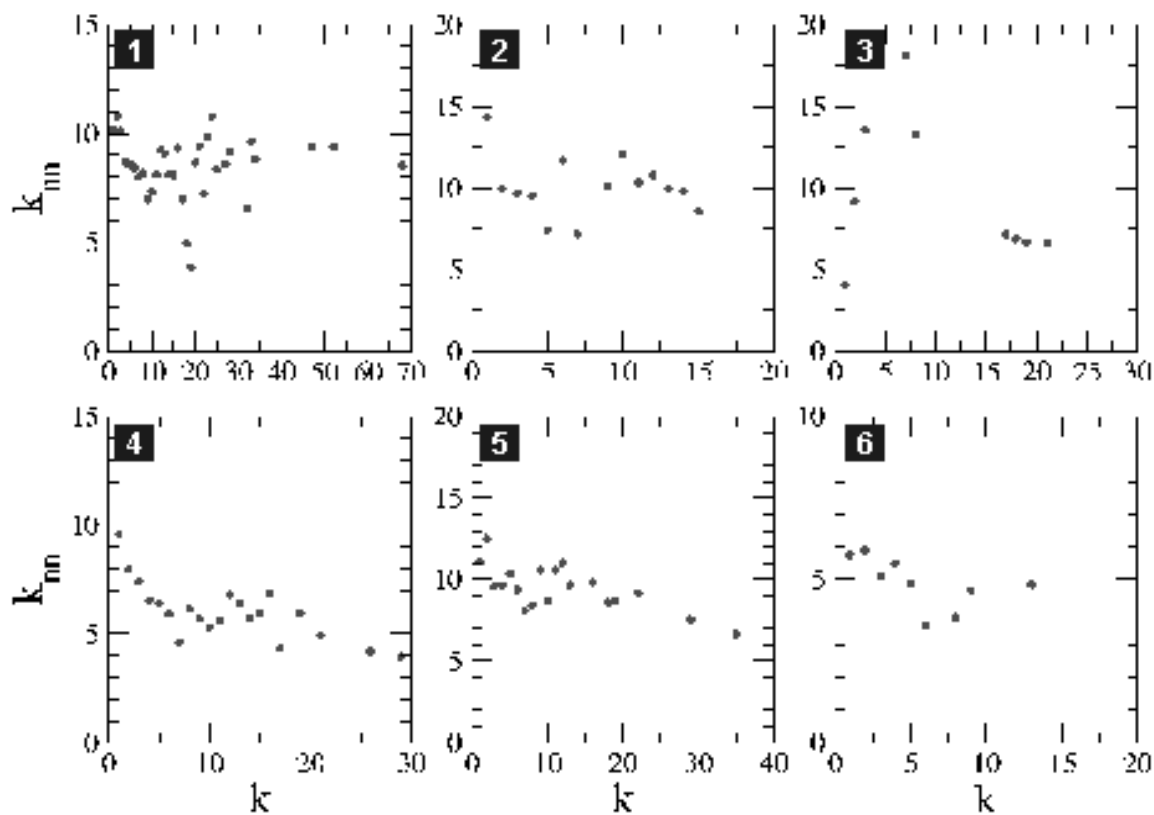

**Fig. 7**

A tendency to disassortativity emerges in Venezia and Wien in this plot of $k_{nn}(k)$, the average degree of nearest neighbours of vertices with degree $k$. In general, the absence of any clue of assortativity in all six cases, with the exception of San Francisco, differentiates street networks from other non geographic systems.

We now turn to evaluate if the dual graphs of the six urban 1-square mile cases are small worlds. We will show that small-world properties clearly emerge in most cases; however, exceptions should be made for networks with few triangular loops.

**Tab. 2**

Characteristic path lengths $L$ and clustering coefficients $C$ of the dual graphs obtained for the six 1-square mile urban patterns considered. The values obtained are compared with those for random graphs with the same size and number of links.

| Case | $L$ | $L_{rand}$ | $C$ | $C_{rand}$ |
|---|---|---|---|---|
| 1  Ahmedabad | 5.20 | 4.81 | 0.250 | 0.003 |
| 2  Barcelona | 2.68 | 2.31 | 0.124 | 0.120 |
| 3  San Francisco | 2.13 | 1.86 | 0.067 | 0.240 |
| 4  Venezia | 8.36 | 5.20 | 0.174 | 0.004 |
| 5  Wien | 3.48 | 3.44 | 0.175 | 0.025 |
| 6  Walnut Creek | 3.96 | 3.44 | 0.062 | 0.026 |

We calculated the average path length and the clustering coefficient of the six networks, and the results, reported in table 2, are compared with those obtained for random graphs with the same number of nodes and links. The networks have a small average path length, smaller than 6 in all the cases considered, Venezia excepted. This indicates that, on average, any two streets on a 1-square mile are only few streets apart. The average distance is particularly small for the dual graphs obtained from grid-like urban patterns (Barcelona and San Francisco). For instance, in San Francisco, any two streets can be connected in just two steps, i.e. with only one intermediate street. In addition, four of the networks considered, namely Ahmedabad, Venezia, Wien and Walnut Creek, have $C >> C_{rand}$, and are therefore small-worlds. Conversely, Barcelona has a clustering coefficient $C$ of the same size of $C_{rand}$, while San Francisco has a clustering coefficient which is even much smaller than $C_{rand}$. This is due to the fact that, as originally defined, the clustering coefficient $C$ is a measure that is related only to the number of triangles present in the network (see section 3.4). In the case of San Francisco such a number is extremely small because of the grid-like structure of the city. As an example consider, for instance, that a city with a perfect square-lattice structure would be mapped into a dual graph with no triangles at all. Similar results can be obtained by computing the network global and local efficiency. All networks result efficient both at the global and local level with the exception of Barcelona and San Francisco, two cases in which, the absence of triangles, affects the value of $E_{loc}$.

These findings, on one hand, imply that the dual networks of Barcelona and San Francisco are not small-worlds, at least according to the usual definition (Watts and Strogatz, 1998; Latora and Marchiori, 2001), since they do not exhibit local clustering as evidenced by the values of $C$ and $E_{loc}$. On the other hand, what we have found means that if we want to better capture and measure the local properties of a network we may need a better definition of $C$ (Jiang and Claramunt, 2004) and $E_{loc}$ (Crucitti et al. 2004), especially for such systems with a small number of triangles.

**Tab. 3**

Global and local efficiency of the dual graphs obtained for the six 1-square mile urban patterns considered. The values obtained are compared with those for random graphs with the same size and number of links.

| Case | $E_{glob}$ | $(E_{glob})_{rand}$ | $E_{loc}$ | $(E_{loc})_{rand}$ |
|---|---|---|---|---|
| 1 Ahmedabad | 0.21 | 0.21 | 0.281 | 0.003 |
| 2 Barcelona | 0.45 | 0.49 | 0.144 | 0.154 |
| 3 San Francisco | 0.57 | 0.60 | 0.070 | 0.400 |
| 4 Venezia | 0.15 | 0.18 | 0.191 | 0.004 |
| 5 Wien | 0.33 | 0.32 | 0.206 | 0.026 |
| 6 Walnut Creek | 0.30 | 0.25 | 0.067 | 0.026 |

Concerning the global properties of the network, we have implemented the procedure proposed by Csanyi and Szendroi (Csanyi and Szendroi, 2004) to assess for the validity of the small-world scaling. Namely, we have calculated $\langle N(r) \rangle$, the average number of nodes of the graph that can be reached from a generic starting node in at most $r$ steps. In fig. 8 we report the results obtained for the two largest cities, Ahmedabad and Venezia. Although the size of the networks does not allow to draw a definitive conclusion, we have plotted the results in figure in a log-log scale because the fractal scaling $\langle N(r) \rangle \sim r^d$ seems to be better verified than the small-world scaling $\langle N(r) \rangle \sim e^{\alpha r}$. This finding confirmed also in networks with a larger number of nodes, it could indicate that the dual graphs still retain some of the geographical constraints of the primal graphs.

As a final step we have investigated the distribution of the node clustering coefficients $C_i$. The clustering coefficient $C_i$ of node $i$ (see section 3.4) has a twofold meaning: on one hand, it tells how cohesive is the cluster of $i$'s first neighbours in terms of their reciprocal relationships; on the other hand, it expresses how critical is $i$ to achieve a direct relationship among all its first neighbours. The latter meaning is specifically inherent the urban case, for it embeds the relevance of one street in terms of its ability to provide a direct link among all the intersecting "secondary" streets: the lower $C_j$, the higher the relevance (or "criticality"). In particular, we have focused our attention on $C(k)$, the average clustering coefficient restricted to classes of vertices of degree $k$.

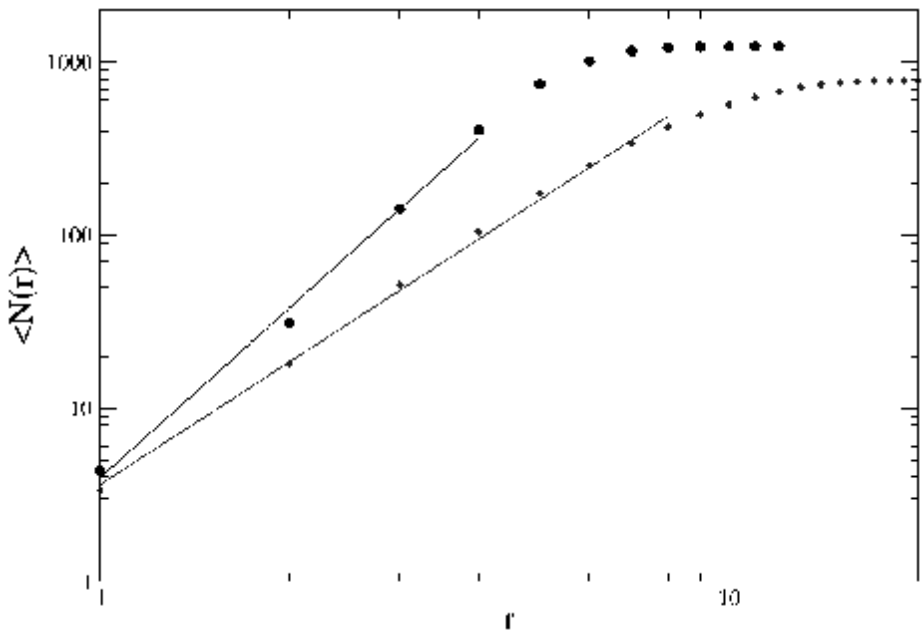

**Fig. 8**

Average number of nodes that can be reached from a generic starting node in at most $r$ steps as a function of r. We report the results for the two largest graphs, namely Ahmedabad (circles) and Venezia (diamonds). For small values of r the two curves are better fitted by power laws (reported as straight lines) rather than exponentials.

The results are reported in figure 9. The same principle of hierarchy previously observed in the degree distributions and correlations, can now be found in the distribution of $C(k)$. In fact, in most of the cases, *C(k)* exhibits a power law decay as a function of *k* (see continuous line in fig. 9) i.e. a hierarchy with low degree vertices belonging to well interconnected communities and hubs connecting many vertices not directly connected between each other. As such, the figure shows a principle of hierarchy emerging – at different "speeds" for the different cities – with the growth of the number of intersections per street: streets with a higher number of intersections tend to be more critical to the local connectivity of their neighbourhood.

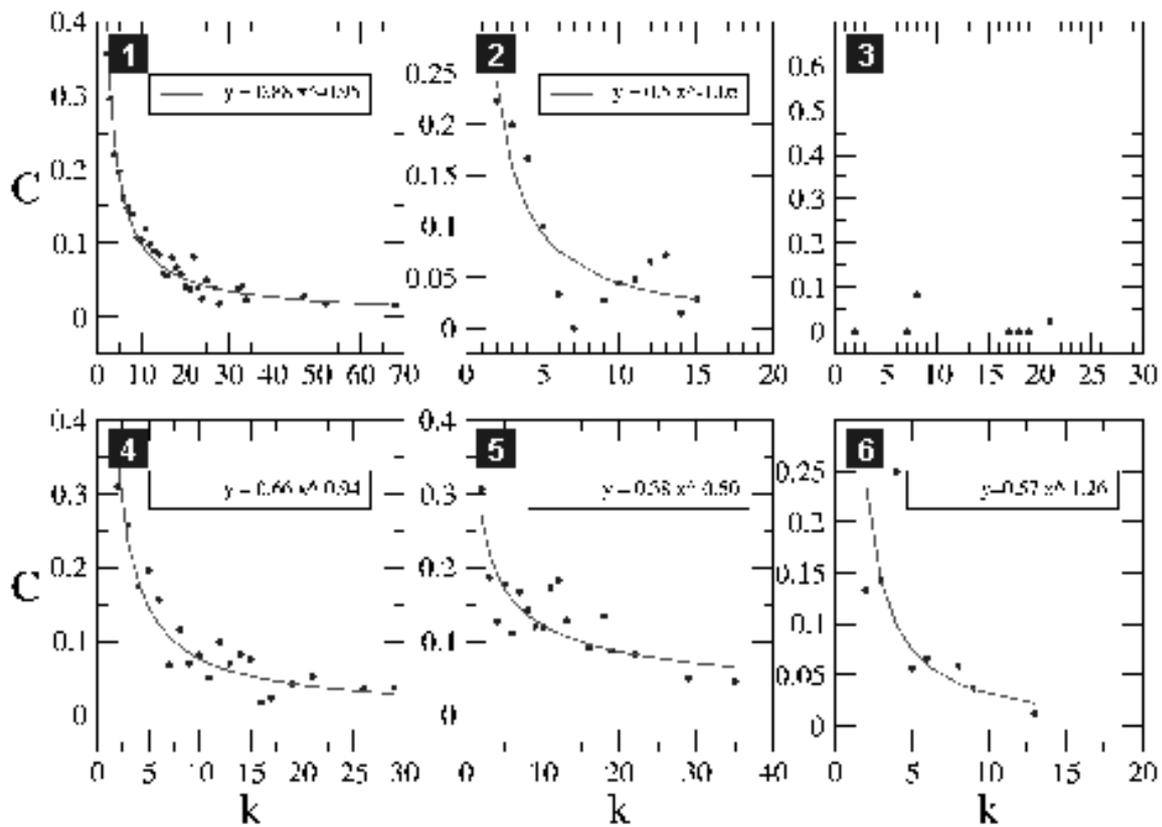

**Fig. 9**

Average clustering coefficient restricted to classes of vertices of degree *k* for the six graphs considered (same order as in figure 5). As for the degree (number of intersected streets), also *C(k)* is distributed according to a power-law. The continuous lines in figures are the fit obtained with a power law.

## 4. Conclusions

The ICN (Intersection Continuity Negotiation) generalization model has been proposed in order to allow the passage from a primal representation of urban street networks to a dual one, where streets are nodes and intersections are edges. Like other models, ICN leads to a loss of any reference to geographic distance, but unlike other models it is purely spatial.

Power law behaviors have been found especially evident in urban street networks of significant size, with reference to the degree distribution and the average clustering coefficient restricted to classes of vertices of degree $k$. By considering the average degree of nearest neighbours of vertices with degree $k$, a tendency to degree disassortativity emerges in Venezia and Wien. However, the absence of any clue of assortativity in all cases differentiates urban street networks, in this dual representation, from other non-geographic systems.

Small world properties have been found emerging as a general rule throughout all cases, with the exception of networks characterised by a low number of triangular loops. However, a definitive conclusion about the small-world scaling should wait further investigations of larger datasets.

Along with these similarities, striking differences have been detected across cases in terms of their simple size: it is amazing how different can urban networks be within the same amount of territorial surface (one square mile). Beside the gap in terms of number of nodes and edges (streets and intersections), smaller networks do exhibit a less obvious though much more relevant feature: in these smaller, less fine-grained cases, like Walnut Creek, Barcelona and San Francisco, it seems that there is simply not enough city in one square mile to make any sense in terms of structural order, i.e. both in terms of power law degree distribution and small-world properties. This latter achievement offers a new argument to the long-term debate about density and sustainability in urban planning.